\documentclass[11pt,a4paper]{article}
\pdfoutput=1
\usepackage{jheppubnohead}
\usepackage[utf8]{inputenc}
\usepackage{amsmath}
\usepackage{epsfig}
\usepackage{graphicx}
\usepackage{amssymb}
\usepackage{tabularx}
\usepackage[normalem]{ulem} 
\usepackage{booktabs} 
\usepackage{subfig}

\providecommand{\proarrow}[0]{\rightarrow}

\providecommand{\dif}[0]{\mathrm{d}}

\providecommand{\proname}[2]{#1 \proarrow #2}

\providecommand{\abs}[1]{\left\lvert #1 \right\rvert}

\providecommand{\abss}[1]{\left\lvert #1 \right\rvert^2}

\providecommand{\mire}[1]{{\rm Re} \left[ #1 \right]}
\providecommand{\miim}[1]{{\rm Im} \left[ #1 \right]}

\providecommand{\order}[1]{{\cal O} \left( #1 \right)}
\providecommand{\torder}[1]{{\cal O} \bigl( #1 \bigr)}

\providecommand{\cpm}[0]{\mathcal{M}}

\providecommand{\pmed}[0]{p_0}
\providecommand{\mmed}[0]{\tilde{M}}
\providecommand{\gmed}[0]{\tilde{\Gamma}}
\providecommand{\emed}[0]{E_0}
\providecommand{\treal}[0]{\theta'_{\rm R}}
\providecommand{\tim}[0]{\theta'_{\rm I}}
\providecommand{\rf}[0]{r}
\providecommand{\pra}[0]{\tilde{M}_1}
\providecommand{\prb}[0]{\tilde{M}_2}
\providecommand{\pia}[0]{\tilde{\Gamma}_1}
\providecommand{\pib}[0]{\tilde{\Gamma}_2}

\newcommand{\be}{\begin{equation}}
\newcommand{\ee}{\end{equation}}
\newcommand{\bea}{\begin{eqnarray}}
\newcommand{\eea}{\end{eqnarray}}

\title{CP violation in mixing and oscillations for leptogenesis~II: the highly degenerate case}
\author[]{J.~Racker}

\affiliation[]{Instituto de Astronom\'{\i}a Te\'orica y Experimental (IATE),  Universidad Nacional
de C\'ordoba (UNC)~- Consejo Nacional de Investigaciones Cient\'{\i}ficas y T\'ecnicas
(CONICET) \\ Laprida 854, X5000BGR,  C\'ordoba, Argentina.\\
Observatorio Astron\'omico de C\'ordoba (OAC), Universidad Nacional de C\'ordoba (UNC) \\ Laprida 854, X5000BGR, C\'ordoba, Argentina.
}

\emailAdd{jracker@unc.edu.ar}

\abstract{We extend to the highly degenerate case a recent approach for analyzing the sources of CP violation in baryogenesis models with quasi-degenerate neutrinos. In this approach an expansion of the resummed propagator around the poles is plugged into a quantum field theory model of neutrino oscillations and a source term for the time evolution of the lepton asymmetry is built directly from the probabilities of lepton number violating processes involving only stable particles. This allows for a transparent consideration of unitarity requirements. The source term has contributions that can be identified with CP violation from mixing, oscillations and interference between both. For the highly degenerate case, i.e. when the mass difference between two neutrinos is similar or smaller than their decay widths, we find that in general the mixing and oscillation terms contribute with opposite signs to the generation of lepton asymmetry and the contribution of the interference term is typically very relevant and crucial to ensure unitarity is satisfied. Moreover, the expressions we obtain are finite in the double degenerate limit of equal masses and couplings. The calculations are done in a simple scalar toy model.}
\begin{document}

\maketitle
\section{Introduction}

Particles with the same quantum numbers and nearly degenerate in mass provide ways for low scale baryogenesis. Notably, in the type I seesaw model it is possible to have leptogenesis in the freeze-out of Majorana neutrinos with $\order{1~{\rm TeV}}$ masses, or in the freeze-in of much lighter neutrinos, also known as resonant leptogenesis ~\cite{pilaftsis03} and baryogenesis via neutrino oscillations (or ARS leptogenesis)~\cite{akhmedov98, asaka05}, respectively. Both have been studied carefully and with different formalisms (see~\cite{Dev:2017wwc} and~\cite{Drewes:2017zyw} for  comprehensive reviews and references). In the last few years there have also been works addressing the case of neutrinos in the intermediate mass range of several tens to hundreds of GeV, which requires a careful implementation of the helicity degree of freedom in the transport equations~\cite{Eijima:2017anv, Ghiglieri:2017gjz, Hambye:2017elz, Ghiglieri:2017csp, Eijima:2018qke, Abada:2018oly, Ghiglieri:2018wbs, Klaric:2020lov, Klaric:2021cpi, Drewes:2021nqr}.  

In these models the CP even phase arises from absorptive parts of loop amplitudes or from oscillating phases due to the coherent propagation of different mass eigenstates. The interplay of these sources of CP violation has been analyzed in detail under different formalisms and approximations in~\cite{Dev:2014laa, Dev:2015wpa, Dev:2014wsa, Kartavtsev15} (see also~\cite{Fidler:2011yq, garbrecht11, garny11, Garbrecht:2014aga}).  In~\cite{Dev:2014laa, Dev:2015wpa} a fully flavor-covariant set of transport equations was derived following a semiclassical approach, while the analysis of~\cite{Dev:2014wsa} and~\cite{Kartavtsev15} are based on the Kadanoff-Baym formalism of non-equilibrium thermal field theory. Although several of the conclusions in these works are compatible, including that the oscillation and mixing sources can contribute additively to the final asymmetry, an interference term between mixing and oscillations was found in~\cite{Kartavtsev15} while not in~\cite{Dev:2014laa, Dev:2015wpa, Dev:2014wsa}. 

Another look at this subject, based on an effective Hamiltonian formalism, was taken long ago in~\cite{Liu:1993ds, covi96II}. Those works elucidated various issues, but a complete implementation of the method to baryogenesis remained open. Motivated by the  simplicity and transparency of this formalism, we followed a quantum field theory approach in~\cite{Racker20}, that up to some point and under certain approximations, can be matched to the effective Hamiltonian formalism. In this approach the renormalized propagator is used in a quantum field theory model of neutrino oscillations in order to obtain probabilities of lepton number violating processes involving only stable particles in the initial and final states. Then a source term for the evolution of the lepton asymmetry is derived from a suitable time integration over the history of the system. Given that this source does not involve processes with -unstable- neutrinos in the initial or final states, no count of neutrino number densities and no subtraction of real intermediate states must be performed, so that unitarity is satisfied in a transparent way. 

The source term obtained in~\cite{Racker20} at first order in the decay widths over the mass splitting has contributions that can be identified with CP violation from mixing, oscillations and interference between both. In equilibrium the terms coming from CP violation in mixing and oscillations cancel, yielding a null source as required by unitarity and CPT invariance. Moreover, in the limit of large mass splittings, this source tends to the one of the standard classical Boltzmann equations appropriate for non-oscillating neutrinos. Therefore, the findings of~\cite{Racker20} do not seem to agree with those of~\cite{Dev:2014laa, Dev:2015wpa, Dev:2014wsa, Kartavtsev15}
regarding the relative sign of the contributions from mixing and oscillations.  

Even more recently, another set of non-equilibrium quantum transport equations for flavor-mixing fermions was derived in~\cite{Jukkala:2021cys} using the Schwinger–Keldysh closed time path formalism and applied to resonant leptogenesis. Although the scope of the equations in~\cite{Jukkala:2021cys} is much bigger than the one of~\cite{Racker20}, it is interesting to note that the results of~\cite{Jukkala:2021cys} also do not support some of the findings of~\cite{Dev:2014laa, Dev:2015wpa, Dev:2014wsa, Kartavtsev15} related to the mixing and oscillation terms and, like in~\cite{Racker20}, the lepton asymmetry in~\cite{Jukkala:2021cys} converges to the usual Boltzmann result in the limit of large mass splittings.

All in all, what seems clear is that the subject of CP violation in leptogenesis with quasi-degenerate neutrinos is not trivial and therefore different looks at the problem can be instructive. With this motivation, here we extend the approach introduced in~\cite{Racker20} to the more involved highly degenerate case, with mass splittings similar or smaller than the decay widths. The work is organized as follows: In section~\ref{sec:toymodel} we calculate a time dependent CP asymmetry between lepton number violating processes involving only stable states, by performing an expansion around the poles of the resummed one-loop propagator and using this expansion in a quantum field theory model for oscillations. Next, in section~\ref{sec:st}, the CP asymmetry is properly integrated over time to get a source term for the generation of lepton asymmetry, which does not involve any count of neutrino number densities. After analyzing the main properties of this source, we summarize the main results and comment on possible directions for future work in section~\ref{sec:con}.

\section{CP asymmetry}
\label{sec:toymodel}
We extend the analysis of~\cite{Racker20} to the highly degenerate case using the same  scalar toy model, which has also been used in several of the references mentioned in the introduction. There is one complex and two real scalar fields, denoted by $b$ and $\psi_i$ ($i=1,2$), respectively. In a basis where the mass matrix of the real scalars is diagonal,
the Lagrangian is given by
\begin{equation}
\label{equation_lag}
\mathcal{L} = \frac{1}{2} \partial^\mu \psi_i \, \partial_\mu \psi_i - \frac{1}{2} \psi_i M_i^2 \psi_i +  \partial^\mu \bar b \, \partial_\mu b - m^2 \, \bar b b - \frac{h_i}{2} \psi_i \, b b - \frac{h_i^*}{2} \psi_i \, \bar b \bar b - \frac{\lambda}{2\cdot 2}(\bar b b)^2 \; .
\end{equation}

The $b$-particles will subsequently be called ``leptons'', since they play in this toy model the analogous role that leptons play in standard leptogenesis, and for simplicity their mass $m$ will be neglected. The lepton charge is broken by the cubic Yukawa interaction terms involving the $\psi_i$, to be called ``neutrinos'' in what follows. The last term is a quartic interaction which does not change lepton number but might be used as a way to localize the leptons and satisfy the conditions to have oscillations~\cite{Beuthe01}, but we will not make explicit use of it. 

The approach to obtain the source term will be the same as in~\cite{Racker20}: First an expansion around the complex poles of the resummed one-loop propagator is performed according to~\cite{fuchs16}, then the amplitudes of lepton number violating processes obtained from this expansion are used in a quantum field theory model for oscillations following~\cite{Beuthe01}, and finally the resulting time dependent probabilities of lepton number violating processes are integrated over time as in~\cite{Racker20} to get the source term. In this section we compute the time dependent CP asymmetry and in the following the source term. Compared to~\cite{Racker20}, care must be taken in several expansions given that we will let the mass difference $M_2 - M_1$, which appears in some denominators, be of the order of, or smaller than, the decay widths.

The one-loop renormalized inverse propagator matrix ${\bf G}^{-1}$ is given by 
 \begin{equation}
 \label{equation_Gdef}
 i {\bf G}^{-1}(p^2) = p^2 {\bf 1} - {\bf M^2}(p^2),
 \end{equation}
 with
 \begin{equation}
 {\bf M^2}(p^2)= 
\begin{pmatrix}
M_1^2 + \Sigma_{11}(p^2)  & \Sigma_{12}(p^2) \\
\Sigma_{21}(p^2)  & M_2^2 + \Sigma_{22}(p^2) 
\end{pmatrix},
\end{equation}
and
\begin{eqnarray}
\Sigma_{ii}(p^2)&=&\frac{\abss{h_i}}{(4\pi)^2} \left[1 + \ln \frac{p^2}{M_i^2} - \frac{p^2}{M_i^2} - i \pi \theta(p^2)\right], \\
\label{equation_sigma12}
\Sigma_{12}(p^2)&=&\Sigma_{21}(p^2)=\frac{\mire{h_1^* h_2}}{(4\pi)^2} \left[\frac{M_2^2 \ln \frac{p^2}{M_1^2} - M_1^2 \ln \frac{p^2}{M_2^2} - p^2 \ln \frac{M_2^2}{M_1^2}}{M_2^2-M_1^2} - i \pi \theta(p^2)\right] .
\end{eqnarray}
 These expressions will always be evaluated for $p^2 > 0$ and therefore the step function $\theta$ will be omitted henceforth. We have used the following renormalization conditions:
 \begin{equation*}
 \mire{\Sigma_{ii}(M_i^2)} = \mire{\Sigma_{12}(M_i^2)} = \frac{\dif \Sigma_{ii}}{\dif p^2}\Bigr\rvert_{p^2=M_i^2} = 0, \quad {\rm for} \; i=1,2\, . \\
\end{equation*}
In the highly degenerate case it might be important to consider the resummation of multi-loop diagrams, however this is out of the goal of this paper and instead we will show that consistent results (finite and complying with unitarity requirements) can be obtained from the one-loop resummed propagator.
Next we will make an expansion of the propagator around the two poles.
 Following~\cite{fuchs16} (see eq. 4.18) we write:
\begin{equation}
\label{equation_gaprox}
{\bf G} \simeq {\bf Z}^{T} \, {\bf \Delta}^{\rm BW} \, {\bf Z} \, ,
\end{equation}
with
\begin{equation}
\label{equation_bw}
{\bf \Delta}^{\rm BW} = i
\begin{pmatrix}
(p^2 - \cpm_{a}^2)^{-1} & 0 \\
0 & (p^2 - \cpm_{b}^2)^{-1} 
\end{pmatrix} ,
\end{equation}
$\cpm_{a,b}^2$ the poles of the propagator, and
\begin{equation}
\label{equation_zmatrix}
{\bf Z} = 
\begin{pmatrix}
\sqrt{Z_1} &  \sqrt{Z_1} Z_{12}\\
\sqrt{Z_2} Z_{21} & \sqrt{Z_2}
\end{pmatrix},
\end{equation}
where
\begin{eqnarray}
\label{equation_zelements}
Z_{12} &=& \frac{{\bf G}_{12}(\cpm_{a}^2)}{{\bf G}_{11}(\cpm_{a}^2)} = \frac{\Sigma_{12}(\cpm_{a}^2)}{\cpm_{a}^2  - M_2^2 - \Sigma_{22}(\cpm_{a}^2)} \, ,  \notag \\
Z_{21} &=& \frac{{\bf G}_{21}(\cpm_{b}^2)}{{\bf G}_{22}(\cpm_{b}^2)} = \frac{\Sigma_{12}(\cpm_{b}^2)}{\cpm_{b}^2  - M_1^2 - \Sigma_{11}(\cpm_{b}^2)} \, , \notag \\
Z_1 &=& \frac{1}{\frac{\partial}{\partial p^2} \frac{i}{{\bf G}_{11} (p^2)}} \Biggr\rvert_{p^2 = \cpm_{a}^2} = \frac{1}{1-\Sigma_{11}^{\rm eff \, '}(\cpm_{a}^2)} \, , \notag \\
Z_2 &=& \frac{1}{\frac{\partial}{\partial p^2} \frac{i}{{\bf G}_{22} (p^2)}} \Biggr\rvert_{p^2 = \cpm_{b}^2} =\frac{1}{1-\Sigma_{22}^{\rm eff \, '}(\cpm_{b}^2)}\, . 
\end{eqnarray}
Moreover, the quantity $\Sigma_{ii}^{\rm eff} (p^2)$ is defined by the identity
\begin{equation}
{\bf G}_{ii} (p^2) = \frac{i}{p^2 - M_i^2 - \Sigma_{ii}^{\rm eff} (p^2)} \, .
\end{equation}
The complex poles $\cpm_{a,b}^2$ are given by the roots of the determinant of ${\bf G}^{-1}$ and each of them satisfies
\begin{equation}
p^2 - M_i^2 - \Sigma_{ii}^{\rm eff} (p^2)\bigr\rvert_{p^2 = \cpm_{a,b}^2} = 0
\end{equation}
for both, $i=1,2$, so that any assignment of the labels $a$ and $b$ to the solutions of these equations is possible (although for numerical purposes one choice might be more convenient than the other~\cite{fuchs16}). At $\torder{h^2}$ (with $h$ representing any of the Yukawa couplings), the poles are equal to
\begin{equation*}
\cpm_a^2 = M_1^2 - i M_1 \Gamma_1 + \order{h^4} \quad {\rm and} \quad
\cpm_b^2 = M_2^2 - i M_2 \Gamma_2 + \order{h^4},
\end{equation*}
where $\Gamma_i \equiv \tfrac{\abss{h_i}}{16 \pi M_i}$.
These approximations are good enough for $\abs{h_i h_j}/(4 \pi)^2 \ll \epsilon$, with $\epsilon \equiv M_2^2 - M_1^2$, as assumed in our previous work~\cite{Racker20}. However, for $\epsilon \lesssim \abs{h_i h_j}/(4 \pi)^2$
 it is necessary to go beyond $\torder{h^2}$ in the calculation of the roots of $|{\bf G}^{-1}|$, due to the presence of $\epsilon$ in some denominators. Then we take
\begin{eqnarray}
\label{equation_ma}
\cpm_a^2 = M_1^2 - i M_1 \Gamma_1 \frac{1+\rf}{2} - i M_2 \Gamma_2 \frac{1-\rf}{2} + \epsilon \frac{1-\rf}{2} , \notag\\
\cpm_b^2 = M_2^2 - i M_2 \Gamma_2 \frac{1+\rf}{2} - i M_1 \Gamma_1 \frac{1-\rf}{2} - \epsilon \frac{1-\rf}{2},
\end{eqnarray} 
with
\begin{equation}
\label{equation_defr}
\rf \equiv \sqrt{1-\left[ \frac{2 \mire{h_1^* h_2}}{16 \pi \epsilon + i \left(\abss{h_1} - \abss{h_2} \right)} \right]^2} \, ,
\end{equation}
for which it can be verified that
\begin{equation}
|{\bf G}^{-1} (\cpm_{a,b}^2)| = 0 + \order{h^8, h^6 \epsilon}\, .
\end{equation}
In addition, the elements of the ${\bf Z}$ matrix defined in eqs.~\eqref{equation_zelements} are equal to
\begin{eqnarray}
\label{equation_zelements2}
Z_{12} &=& -\theta' + \order{h^4} = - Z_{21} + \order{h^4} \, ,\notag \\ 
Z_1 &=& \frac{1+\rf}{2 \rf}  + \order{h^4} = Z_2  + \order{h^4}\, , 
\end{eqnarray}
where
\begin{equation}
\label{equation_thetaprime}
- \theta' \equiv  \frac{i \, \mire{h_1^* h_2}/(16 \pi)}{\epsilon+ i \left(M_1 \Gamma_1 - M_2 \Gamma_2 \right)} \, \frac{2}{1+\rf} = \sqrt{\frac{\rf - 1}{\rf + 1}} 
\end{equation}
and the $\order{h^4}$ terms are finite in the limit $\epsilon \to 0$.
Note that we are attaching a prime to $\theta$ in order to distinguish it from the analog quantity defined in~\cite{Racker20}. In terms of $\theta'$, the poles in eqs.~\eqref{equation_ma} read
\begin{eqnarray}
\cpm_a^2 &=& M_1^2 - i M_1 \Gamma_1 + i \, \frac{\mire{h_1^* h_2}}{16 \pi} \, \theta' \, , \notag\\
\cpm_b^2 &=& M_2^2 - i M_2 \Gamma_2 - i \, \frac{\mire{h_1^* h_2}}{16 \pi} \, \theta' \, ,
\end{eqnarray}
so that the real and imaginary parts are given by
\begin{eqnarray}
\label{equation_cp2}
\cpm_a^2 &=& \left(M_1^2 - \frac{\mire{h_1^* h_2}}{16 \pi} \, \miim{\theta'} \right)  - i \, \left(M_1 \Gamma_1 - \frac{\mire{h_1^* h_2}}{16 \pi} \, \mire{\theta'} \right) \, , \notag \\
 \cpm_b^2 &=& \left(M_2^2 + \frac{\mire{h_1^* h_2}}{16 \pi} \, \miim{\theta'} \right)  - i \, \left(M_2 \Gamma_2 + \frac{\mire{h_1^* h_2}}{16 \pi} \, \mire{\theta'} \right) \, .
\end{eqnarray}
It can be checked that the imaginary parts of $\cpm_{a,b}^2$ are always negative.

From the expansion of the propagator around the complex poles specified in eqs.~\eqref{equation_gaprox}-\eqref{equation_zmatrix} and~\eqref{equation_zelements2}, the invariant matrix elements for the lepton number violating processes become 
\begin{eqnarray*}
-\mathcal{M}(\proname{\bar b \bar b}{b b}) & = & - i \sum_{j,k} h_j^* \, {\bf G}_{j k} \, h_k^* \\ &=& Z_1 \left( h_1^{* 2} - 2 h_1^* h_2^* \, \theta' + h_2^{* 2}\, \theta'^{\,2} \right) \Delta_1 +   Z_1 \left( h_2^{* 2} + 2 h_1^* h_2^* \,\theta' + h_1^{* 2} \,\theta'^{\,2}\right) \Delta_2 \, , \\
-\mathcal{M}(\proname{b b}{\bar b \bar b}) & = & -i \sum_{j,k}  h_j {\bf G}_{j k}  h_k \\ &=& Z_1 \left( h_1^2 - 2 h_1 h_2 \,\theta' + h_2^2 \,\theta'^{\,2} \right) \Delta_1 +   Z_1 \left( h_2^2 + 2 h_1 h_2 \,\theta' + h_1^2 \,\theta'^{\,2}\right) \Delta_2  , 
\end{eqnarray*}
with
\begin{equation*}
\Delta_{1\,(2)} \equiv \frac{1}{p^2 - \cpm_{a \, (b)}^2} \, .
\end{equation*}
These amplitudes can be used in a quantum field theory model of oscillations. We will consider an external wave packet model~\cite{1963AnPhy22, PhysRevD.48.4310} following the detailed review and analysis of~\cite{Beuthe01}. In this model the initial and final states of a given process are described by localized wave packets. Assuming that the factors related to coherence and localization, which could destroy oscillations, can be neglected, the following expressions for the probabilities of the lepton number violating processes are obtained:
\begin{eqnarray}
\label{equation_AL}
\abss{A}(L)  &=&  N \left|\left( h_1^{* 2} - 2 h_1^* h_2^* \,\theta' + h_2^{* 2} \,\theta'^{\,2}\right)  e^{-i \left(\pra - i \frac{\pia}{2}\right) \frac{\mmed L}{\pmed}} \; + \notag \right.\\ & & \quad \; \left. \left( h_2^{* 2} + 2 h_1^* h_2^* \,\theta' + h_1^{* 2} \,\theta'^{\,2} \right)    e^{-i \left(\prb - i \frac{\pib}{2}\right) \frac{\mmed L}{\pmed}}  \right|^2 \, , \notag \\
\abss{\bar A}(L)  &=&  N \left|  \left( h_1^2 - 2 h_1 h_2 \,\theta' + h_2^2\, \theta'^{\,2} \right) e^{-i \left(\pra - i \frac{\pia}{2}\right) \frac{\mmed L}{\pmed}} \; + \notag \right.\\ & & \quad \; \left.   \left( h_2^2 + 2 h_1 h_2 \,\theta' + h_1^2\, \theta'^{\,2} \right) e^{-i \left(\prb - i \frac{\pib}{2}\right) \frac{\mmed L}{\pmed}}  \right|^2 \, .
\end{eqnarray}
Here, to simplify the notation we have defined $A \equiv A(\proname{\bar b \bar b}{b b})$ and $\bar A \equiv A(\proname{b b}{\bar b \bar b})$,  $L$ is the distance between the production and decay of the neutrinos that mediate these processes, and $\pmed$ is the average momentum of the neutrinos. We have integrated over solid angle and the normalization constant $N$ can be determined as in~\cite{Racker20} (see~\cite{Beuthe01} for the case of stable neutrinos). The real quantities $\pra, \prb, \pia$ and $\pib$ are defined from the real and imaginary parts of $\cpm_{a, b}$ via the relations (see eq.~\eqref{equation_cp2}):
\begin{eqnarray}
\pra^2 &=& \left(M_1^2 - \frac{\mire{h_1^* h_2}}{16 \pi} \, \miim{\theta'} \right) , \; \pra \pia = \left(M_1 \Gamma_1 - \frac{\mire{h_1^* h_2}}{16 \pi} \, \mire{\theta'} \right) \, ,\notag \\
\prb^2 &=& \left(M_2^2 + \frac{\mire{h_1^* h_2}}{16 \pi} \, \miim{\theta'} \right)  , \; \prb \pib = \left(M_2 \Gamma_2 + \frac{\mire{h_1^* h_2}}{16 \pi} \, \mire{\theta'} \right) \, ,
\end{eqnarray}
and $\mmed \equiv (\pra+\prb)/2$.

The expressions~\eqref{equation_AL} for the probabilities are valid up to first order in $(M_2^2-M_1^2)/(2 \pmed^2)$ and can be matched to an effective Hamiltonian approach within the approximations we have made~\cite{Beuthe01}.
For the following discussion it will be more convenient to change from distance $L$ to time $t$ via the relation $\tfrac{\mmed L}{\pmed}=\tfrac{t}{\gamma}$, with $\gamma \equiv \emed/\mmed$ the Lorentz factor and $\emed$ the average energy (i.e. $\tfrac{\mmed L}{\pmed}$ is the classical proper time of propagation).

Next we compute the CP asymmetry $\abss{A} - \abss{\bar A}$ from eqs.~\eqref{equation_AL}, noticing that there are two different types of CP even phases: one independent of $L$ (or $t$) in $\theta'$, and an oscillating one in the exponentials  $e^{-i  {\tilde M}_j  t/ \gamma}$. Considering all the interferences and the corresponding source of the CP even relative phases, the CP asymmetry can be written as a sum of contributions from mixing $M$ (involving only $\theta'$), from oscillations $O$ (involving only $e^{-i  {\tilde M}_j  t/ \gamma}$), and interference terms $I$ (involving both $\theta'$ and $e^{-i  {\tilde M}_j  t/ \gamma}$):
\begin{equation}
\label{equation_deltaA}
 \frac{\abss{A}(t) - \abss{\bar A}(t)}{N} = M(t) + O(t) + I(t) \, ,
\end{equation}  
where
\begin{eqnarray}
\label{equation_AM}
M(t) &= & 8 \, \miim{h_1 h_2^*} \tim \left\{ e^{-\pia t/\gamma} \left[\abss{h_1} + \abss{h_2} \abss{\theta'} - 2 \mire{h_1^* h_2} \treal \right]  \right. \notag \\ & & \left. +  e^{-\pib t/\gamma} \left[\abss{h_2} + \abss{h_1} \abss{\theta'} + 2 \mire{h_1^* h_2} \treal \right] \right\}, \\
\label{equation_AO}
O(t) & = & 8 \, \miim{h_1 h_2^*} \mire{h_1^* h_2} \left( 1- \abs{\theta'}^4 \right) \miim{e^{i (\prb - \pra) t/\gamma}} e^{-\gmed t/\gamma}, \\ 
\label{equation_AI}
I(t) & = &  8 \, \miim{h_1 h_2^*} \left( 1+\abss{\theta'} \right) \left\{ \treal \,  \miim{e^{i (\prb - \pra) t/\gamma}} e^{-\gmed t/\gamma} \left(\abss{h_1} - \abss{h_2}\right) \right. \notag \\ & & \left. - \tim \, \mire{e^{i (\prb - \pra) t/\gamma}} e^{-\gmed t/\gamma} \left(\abss{h_1} + \abss{h_2}\right)\right\}.
\end{eqnarray}
Here we have defined $\gmed \equiv (\pia+\pib)/2$, while $\treal$ and $\tim$ denote the real and imaginary parts of $\theta'$, respectively.

\section{Source term and analysis}
\label{sec:st}

The time evolution of the lepton density asymmetry $n_L \equiv n_b - n_{\bar b}$, with $n_b$ ($n_{\bar b}$) the number density of leptons (antileptons), can be obtained from the sum of two terms, 
 \begin{equation}
 \frac{\dif n_L}{\dif t} = S(t) - W(t),
 \end{equation}
 where the source $S(t)$ is the part which may be non-null in the absence of a lepton density asymmetry and $W(t)$ is the so-called washout term. The source can be obtained from a proper integration of the CP asymmetry in eq.~\eqref{equation_deltaA} over the whole history of the system, without resorting to some count of neutrino number densities, as explained in~\cite{Racker20}. For our purposes it is enough to consider a static universe and that all the neutrinos mediating the processes in eqs.~\eqref{equation_AL} have the same average momentum $p_0$, so that momentum integrals are avoided. Moreover, finite density effects will not be included.
Therefore the normalization constant $N$ in eq.~\eqref{equation_deltaA} is the same as in~\cite{Racker20}, $N=1/(32 \pi \emed)^2$, with $\emed$ the average energy, and the source term reads
 \begin{equation}
 \label{equation_source}
 S(T) = 2 \int_0^T \frac{n^{\rm eq}(t)}{(32 \pi \emed)^2} \, \Big[M(T-t) + O(T-t) + I(T-t) \Big] \, \dif t \, .
 \end{equation}
 The time dependent functions $M, O$ and $I$ are given by eqs.~\eqref{equation_AM}-\eqref{equation_AI} and $n^{\rm eq}(t)$ is the equilibrium density of a scalar particle of mass $\mmed$. Although in realistic calculations $n^{\rm eq}(t)$ would be a function of the time dependent temperature, in the examples given below for a static universe we will artificially vary $n^{\rm eq}(t)$ and equilibrium will simply correspond to constancy over time.

\begin{figure}[!t]
\centerline{\protect\hbox{
\epsfig{file=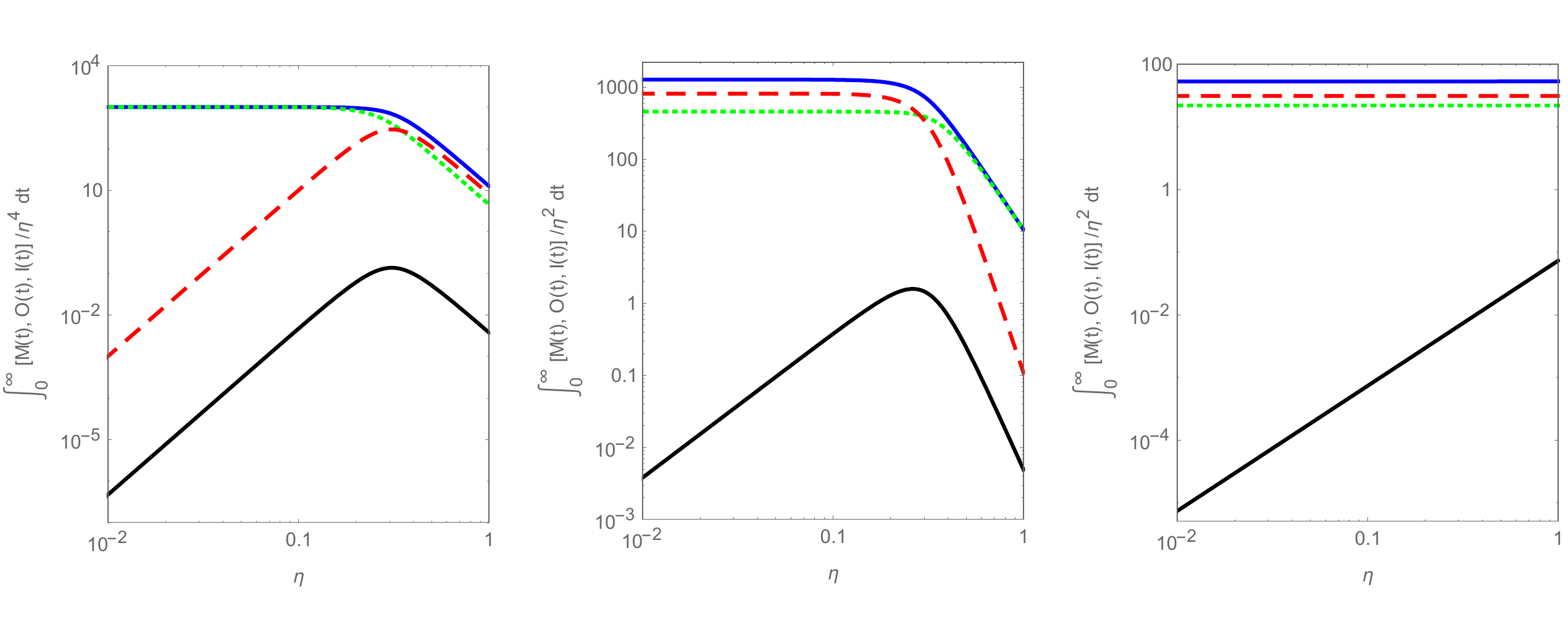,width=0.99\textwidth,angle=0}}} 
\caption[]{Behavior of CP odd quantities for small Yukawa couplings and mass splittings, namely $\abs{\int_0^\infty  M(t) \, \dif t}/(M_1^4 \, \eta^a)$ (solid blue lines), $\abs{\int_0^\infty  O(t) \, \dif t}/(M_1^4 \, \eta^a)$ (dotted green lines), $\abs{\int_0^\infty  I(t) \, \dif t}/(M_1^4 \, \eta^a)$ (dashed red lines), and the CP asymmetry $\abs{\int_0^\infty  \, [\abss{A}(t)-\abss{\bar A}(t)]/N \, \dif t}/(M_1^4 \, \eta^a) =$ $\abs{\int_0^\infty M(t) + O(t) + I(t) \, \dif t}/(M_1^4 \, \eta^a)$ (thick solid black lines), as a function of $\eta$, which parametrizes $h_{1,2}$ and $\epsilon$. Note that we have normalized these quantities to $M_1^4$ to make them dimensionless and to a certain power of $\eta$ so that the mixing and oscillation contributions be constant in the limit $\eta \to 0$ ($a=4$ in the left plot, and $a=2$ in the middle and right plots). For all plots we have chosen $h_1/M_1= h$ and $h_2/M_1=(h/2) e^{i \pi/4}$, while $h=\eta$ and $\epsilon/M_1^2 = \tfrac{1}{10} \tfrac{1}{16 \pi}$ in the left plot, $h=1$ and $\epsilon/M_1^2 = 10 \tfrac{1}{16 \pi} \eta^2$ in the middle plot, and $h=\eta$ and $\epsilon/M_1^2 =  \tfrac{1}{2} \tfrac{1}{16 \pi} \eta^2$ in the right plot. 
} 
\label{fig:0}
\end{figure}

Unitarity and CPT invariance imply that, for a given initial state, $\sum_j \abss{A(\proname{i}{j})} = \sum_j \abss{A(\proname{\bar i}{\bar j})}$ (with the bar denoting CP conjugate states). For our scalar toy model, considering stable asymptotic states with well defined momentum, this implies, to lowest non-trivial order in the couplings (so that only processes with two final particles need to be considered), that 
\begin{equation*}
\abss{A(\proname{b b}{b b})} + \abss{A(\proname{b b}{\bar b \bar b})} = \abss{A(\proname{\bar b \bar b}{\bar b \bar b})} + \abss{A(\proname{\bar b \bar b}{b b})},
\end{equation*}
and therefore
\begin{equation}
\label{equation_unit}
\Delta \abss{A(\proname{\bar b \bar b}{b b})} \equiv \abss{A(\proname{\bar b \bar b}{b b})}  -  \abss{A(\proname{b b}{\bar b \bar b})} = 0 . 
\end{equation}
As already verified in~\cite{Racker20}, this unitarity requirement is satisfied using the exact one-loop resummed propagator $\bf{G}$ given by eqs.~\eqref{equation_Gdef}-\eqref{equation_sigma12}\footnote{There are also one-loop vertex contributions at the same order in the Yukawa couplings, but they cancel independently in eq.~\eqref{equation_unit}, see e.g.~\cite{roulet97}.}. From this fact it is possible to determine the order at which the unitarity condition will be satisfied for a given approximation to $\bf{G}$. 
Note also that in eqs.~\eqref{equation_AL} the states, in particular the final ones, have been taken as wave packets localized in space. Therefore the sum over all possible final states in the unitarity requirement involves an integral over $L$, or equivalently over $t$. Indeed we have verified numerically that $\int_0^\infty  \abss{A}(t) \, \dif t$ equals  $\int_0^\infty  \abss{\bar A}(t) \, \dif t$ up to terms which are higher order in the Yukawa couplings and $\epsilon$ than the individual contributions from mixing, oscillations and interference to the time integral of the CP asymmetry. Specifically, letting $h_{1,2}$ and $\epsilon$ go to zero with some powers of a small parameter $\eta$, i.e. $h_{1,2} \propto h \propto \eta^{a_h}$ and $\epsilon \propto \eta^{a_\epsilon}$, in the limit $\eta \to 0$ the integrals approach zero as
\begin{eqnarray*}
\int_0^\infty M(t), O(t) \, \dif t &\propto&  h^4 \, \frac{1}{\epsilon^2 + h^4} \; \epsilon \, ,\\
\int_0^\infty I(t) \, \dif t &\propto& \left(h^4 \, \frac{1}{\epsilon^2 + h^4}\right)^2 \epsilon \, ,\\
\int_0^\infty \left[\abss{A}(t) - \abss{\bar A}(t)\right] \, \dif t &\propto& \left(h^4 \, \frac{1}{\epsilon^2 + h^4}\right)^2 \epsilon^2\, .
\end{eqnarray*}
This is illustrated in figure~\ref{fig:0} for three cases: $h_{1,2}$ going to zero with $\epsilon$ kept constant, which was the case studied in~\cite{Racker20} ($a_h=1, a_\epsilon=0$, left plot), $\epsilon$ going to zero with $h_{1,2}$ fixed ($a_h=0, a_\epsilon=2$, middle plot), and the three parameters going simultaneously to zero, with $\epsilon \propto \Gamma_{1}$ ($a_h=1, a_\epsilon=2$, right plot). In all cases it is apparent that $\abs{\int_0^\infty M(t) + O(t) + I(t)  \, \dif t}$ is higher order in the small parameters and much lower than $\abs{\int_0^\infty  M(t) \, \dif t} + \abs{\int_0^\infty  O(t) \, \dif t} + \abs{\int_0^\infty  I(t) \, \dif t}$, which makes our approach consistent regarding unitarity. Note that in equilibrium, i.e.~when $n^{\rm eq}(t)$ remains constant for a time period larger than the other time scales (the oscillation period and lifetimes of neutrinos), the unitarity condition ensures that $S(t)$ becomes null. Another related consequence is that the final lepton asymmetry obtained by integrating the source term and neglecting washouts is also null, within the validity of our approximations, if $n^{\rm eq}(t)$ is zero outside a certain window of time. This will be apparent in the figures shown below by the negligible value of the net final lepton asymmetry compared to -at least some of- the individual contributions from mixing, oscillations and interference.

\begin{figure}[!t]
\centerline{\protect\hbox{
\epsfig{file=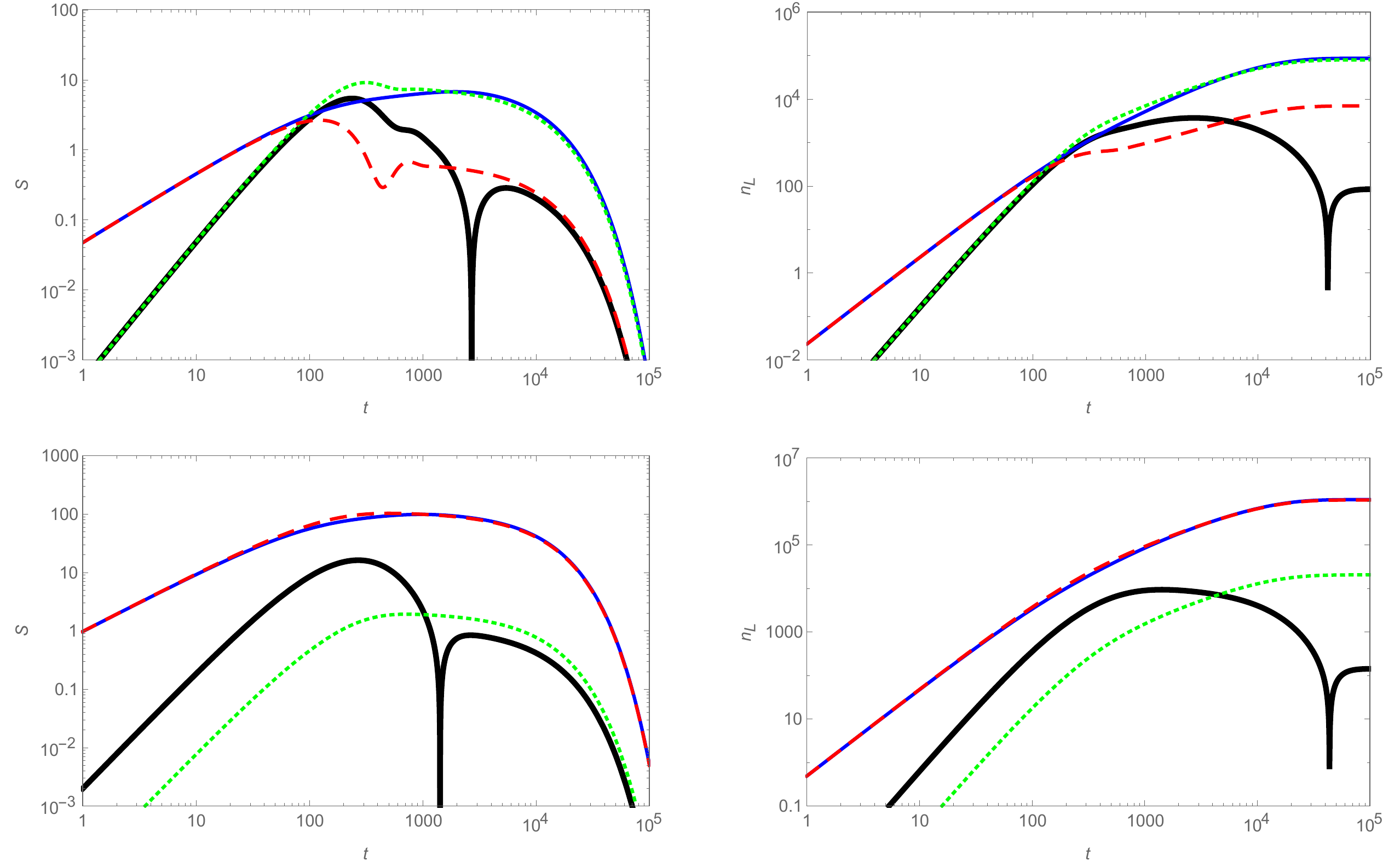,width=0.95\textwidth,angle=0}}} 
\caption[]{Absolute value of the contributions to the source term (left plots) and lepton asymmetry (right plots), as a function of time normalized to $\gamma/M_1$. The solid blue lines represent the mixing contribution (according to eq.~\eqref{equation_AM}), the dotted green lines the oscillation contribution (eq.~\eqref{equation_AO}), the dashed red lines the interference contribution (eq.~\eqref{equation_AI}), and the thick solid black lines give the absolute value of the sum of all contributions, i.e.~the total source term (eq.~\eqref{equation_source}) for the left plots and the net lepton asymmetry for the right ones. In the top plots we have chosen $\Gamma_1/M_1=1/100$, $\Gamma_2/M_1=1/1000$ and $\Delta M=\Gamma_1$, while the corresponding quantities with tilde differ by no more than 10\% in this case. For the bottom plots, $\Gamma_1/M_1=1/100$, $\Gamma_2/M_1=1/130$ and $\Delta M=0.2 \, \Gamma_1$, so that $\pia/M_1\simeq 0.015$, $\pib/M_1 \simeq 0.003$ and $\Delta {\tilde M} \simeq 0.03 \,\pia$. In all cases we have taken $h_1=\abs{h_1}$ and $h_2=\abs{h_2} e^{i \phi}$, with $\phi=\pi/4$. The lepton asymmetry has been obtained integrating only the source term (washouts are not considered). The scale on the vertical axis is not relevant and we have taken, for the purpose of illustration, $n^{\rm eq}(t)=e^{-M_1 \, t/(10000 \, \gamma)}$ (after a change of variables in the integration over time, the factor $M_1/\gamma$ becomes part of the normalization chosen for the lepton asymmetry).} 
\label{fig:1}
\end{figure}

\begin{figure}[!t]
\centerline{\protect\hbox{
\epsfig{file=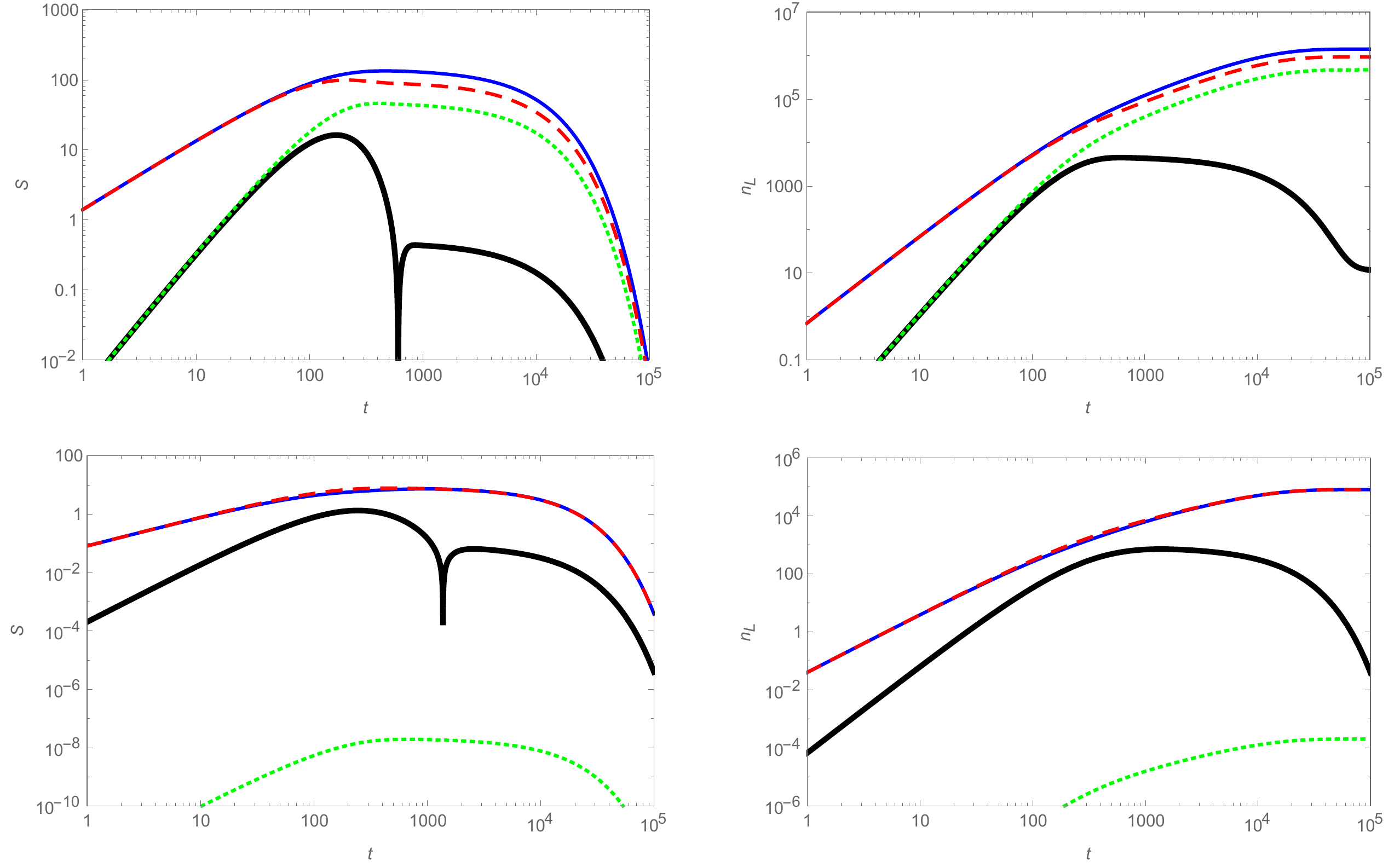,width=0.95\textwidth,angle=0}}} 
\caption[]{
Same as figure~\ref{fig:1} but with another selection of couplings and masses: $\Gamma_1/M_1=\Gamma_2/M_1=1/100$ and $\Delta M= 1 \, (0.01) \, \Gamma_1$ in the top (bottom) plots. Here we have also taken $h_1=\abs{h_1}$ and $h_2=\abs{h_2} e^{i \phi}$, with $\phi=\pi/4$. Therefore in the top plots $\pia/M_1\simeq \pib/M_1 \simeq 1/100$ and $\Delta {\tilde M} \simeq 0.7 \,\pia$, while in the bottom plots $\pia/M_1\simeq 0.003$, $\pib/M_1 \simeq 0.017$ and $\Delta {\tilde M} \simeq 2 \times 10^{-6} \;\pia$.
} 
\label{fig:2}
\end{figure}

To discuss some of the main features of the source term given by eq.~\eqref{equation_source}, we plot in figures~\ref{fig:1} and~\ref{fig:2} the evolution of the different contributions to the source (left plots) and lepton asymmetry (right plots), for different choices of the couplings and masses.  In all cases we have taken $n^{\rm eq}(t)=e^{-M_1 \, t/(10000 \, \gamma)}$, so that the time scale of the evolution of $n^{\rm eq}$ is much larger than the lifetimes of the neutrinos. The lepton asymmetry has been obtained by integrating the source term over time, without considering any washouts. Therefore, as noted above, unitarity requires that the final asymmetry be null. Indeed, this behavior can clearly be seen in the plots, which show that the final value of the lepton asymmetry (solid black lines) is negligible compared to -at least some of- the individual contributions from mixing, oscillations and interference (colored lines). In the top plots of figure~\ref{fig:1} we have chosen neutrinos with lifetimes differing by one order of magnitude. The contribution from the interference term becomes small at late times compared to the contributions from mixing and oscillations, but its role to ensure that the final asymmetry be null is nevertheless important. At small times (compared to the lifetimes and oscillation period), the interference term almost cancels the mixing one, so that the net lepton asymmetry equals the contribution from oscillations. However, we show with an example in the bottom plots of figure~\ref{fig:1} that this might not always be the case. Namely it may happen that at small times the mixing and interference terms partially cancel, but the oscillation contribution remains subdominant, so that the net lepton asymmetry is determined from the former terms and not from oscillations. This behavior at small times can be understood from a Taylor expansion of eqs.~\eqref{equation_AM}-\eqref{equation_AI}. In this second example the oscillation term remains subdominant also at late times, but again unitarity is satisfied due to a cancellation arising from the three contributions.   

The time dependent CP asymmetry we have obtained in the previous section remains finite in the double degenerate limit of equal masses and couplings (see in particular eqs.~\eqref{equation_zelements2}-\eqref{equation_thetaprime}). In order to illustrate the double degenerate limit, we have chosen for figure~\ref{fig:2} $\Gamma_1/M_1=\Gamma_2/M_1=1/100$ and $\Delta M= 1 \, (0.01) \, \Gamma_1$ in the top (bottom) plots. It can be seen that the lepton asymmetry remains finite and decreases in the more degenerate case. Also note that in the bottom plots the oscillation period is much larger than the lifetimes of the neutrinos, therefore the contribution from oscillations becomes negligible. 


\section{Conclusions and outlook}
\label{sec:con}
We have studied the sources of CP violation in a scalar toy model for baryogenesis with highly degenerate neutrinos, i.e. with mass splittings similar or smaller than the decay widths, extending in this way the analysis started in~\cite{Racker20} for milder degeneracies. The approach follows these steps: (1) perform an expansion around the poles of the resummed propagator (see eq.~\eqref{equation_gaprox}), (2) use this expansion in a quantum field theory model for neutrino oscillations to calculate a time dependent CP asymmetry between the probabilities of lepton number violating processes (eq.~\eqref{equation_deltaA}), and (3) this CP asymmetry, which only involves processes with stable initial and final states, must be properly integrated over time to obtain a source term for the evolution of the lepton asymmetry (eq.~\eqref{equation_source}). 

The source term has contributions that can be identified with CP violation from mixing, oscillations and interference between both. The interference term is typically very relevant and crucial to ensure unitarity is satisfied (this is apparent in figure~\ref{fig:0} and also manifests in figure~\ref{fig:1} by the negligible value of the final lepton asymmetry compared to the individual contributions from mixing, oscillations and interference). Moreover, the expressions we obtain are finite in the double degenerate limit of equal masses and couplings, as can be seen in the equations derived in section~\ref{sec:toymodel} and illustrated in figure~\ref{fig:2}. At early times the interference contribution tends to cancel the mixing part, but the net lepton asymmetry might or might not be dominated by the oscillation term, as can be seen from eqs.~\eqref{equation_AM}-\eqref{equation_AI} and the examples in figures~\ref{fig:1} and~\ref{fig:2}. 

In general we find, as in~\cite{Racker20}, that the mixing and oscillation terms contribute with opposite signs to the source, which does not support some of the results of~\cite{Dev:2014laa, Dev:2015wpa, Dev:2014wsa, Kartavtsev15}. We have noted in the introduction that the results of~\cite{Jukkala:2021cys}, based on the Schwinger–Keldysh closed time path formalism, also do not support some of the findings of~\cite{Dev:2014laa, Dev:2015wpa, Dev:2014wsa, Kartavtsev15} related to the mixing and oscillation terms and, like in~\cite{Racker20}, the lepton asymmetry in~\cite{Jukkala:2021cys} converges to the usual Boltzmann result in the limit of large mass splittings. The authors of~\cite{Jukkala:2021cys} suggest that the discrepancy might come from the helicity dependence, noting that the works~\cite{Dev:2014laa, Dev:2015wpa, Dev:2014wsa, Kartavtsev15} are based in a  scalar toy model or a semiclassical approach. However the method described here and in~\cite{Racker20} has been implemented in the same scalar toy model of~\cite{Dev:2014wsa, Kartavtsev15}, suggesting that the reason for the discrepancy may lie somewhere else.

The treatment of CP violation in leptogenesis models with quasi-degenerate neutrinos seems not to be trivial and has actually been discussed over some decades now. Therefore different approaches can be helpful to understand this problem. The one developed here is particularly transparent regarding the key requirements from unitarity and CPT invariance. 
We have performed this study and~\cite{Racker20} in a simple scalar toy model, within a static universe, and for a trivial momentum distribution of the particles. Some of the next steps could be to extend the approach to an expanding universe and spin 1/2 neutrino fields with realistic momentum distributions, as well as to other type of scattering processes, in order to make a closer connection to ARS and resonant leptogenesis, including the intermediate mass regime.

\bibliographystyle{JHEP}
\bibliography{referencias_leptogenesis3}

\providecommand{\href}[2]{#2}\begingroup\raggedright\begin{thebibliography}{10}

\bibitem{pilaftsis03}
A.~Pilaftsis and T.~E. Underwood, {\it {Resonant leptogenesis}},  {\em
  Nucl.Phys.} {\bf B692} (2004) 303--345,
  [\href{http://arxiv.org/abs/hep-ph/0309342}{{\tt hep-ph/0309342}}].

\bibitem{akhmedov98}
E.~K. Akhmedov, V.~Rubakov, and A.~Y. Smirnov, {\it {Baryogenesis via neutrino
  oscillations}},  {\em Phys.Rev.Lett.} {\bf 81} (1998) 1359--1362,
  [\href{http://arxiv.org/abs/hep-ph/9803255}{{\tt hep-ph/9803255}}].

\bibitem{asaka05}
T.~Asaka and M.~Shaposhnikov, {\it {The nuMSM, dark matter and baryon asymmetry
  of the universe}},  {\em Phys.Lett.} {\bf B620} (2005) 17--26,
  [\href{http://arxiv.org/abs/hep-ph/0505013}{{\tt hep-ph/0505013}}].

\bibitem{Dev:2017wwc}
B.~Dev, M.~Garny, J.~Klaric, P.~Millington, and D.~Teresi, {\it {Resonant
  enhancement in leptogenesis}},  {\em Int. J. Mod. Phys. A} {\bf 33} (2018)
  1842003, [\href{http://arxiv.org/abs/1711.02863}{{\tt arXiv:1711.02863}}].

\bibitem{Drewes:2017zyw}
M.~Drewes, B.~Garbrecht, P.~Hern\'andez, M.~Kekic, J.~Lopez-Pavon, J.~Racker,
  N.~Rius, J.~Salvado, and D.~Teresi, {\it {ARS Leptogenesis}},  {\em Int. J.
  Mod. Phys. A} {\bf 33} (2018), no.~05n06 1842002,
  [\href{http://arxiv.org/abs/1711.02862}{{\tt arXiv:1711.02862}}].

\bibitem{Eijima:2017anv}
S.~Eijima and M.~Shaposhnikov, {\it {Fermion number violating effects in low
  scale leptogenesis}},  {\em Phys. Lett. B} {\bf 771} (2017) 288--296,
  [\href{http://arxiv.org/abs/1703.06085}{{\tt arXiv:1703.06085}}].

\bibitem{Ghiglieri:2017gjz}
J.~Ghiglieri and M.~Laine, {\it {GeV-scale hot sterile neutrino oscillations: a
  derivation of evolution equations}},  {\em JHEP} {\bf 05} (2017) 132,
  [\href{http://arxiv.org/abs/1703.06087}{{\tt arXiv:1703.06087}}].

\bibitem{Hambye:2017elz}
T.~Hambye and D.~Teresi, {\it {Baryogenesis from L-violating Higgs-doublet
  decay in the density-matrix formalism}},  {\em Phys. Rev. D} {\bf 96} (2017),
  no.~1 015031, [\href{http://arxiv.org/abs/1705.00016}{{\tt
  arXiv:1705.00016}}].

\bibitem{Ghiglieri:2017csp}
J.~Ghiglieri and M.~Laine, {\it {GeV-scale hot sterile neutrino oscillations: a
  numerical solution}},  {\em JHEP} {\bf 02} (2018) 078,
  [\href{http://arxiv.org/abs/1711.08469}{{\tt arXiv:1711.08469}}].

\bibitem{Eijima:2018qke}
S.~Eijima, M.~Shaposhnikov, and I.~Timiryasov, {\it {Parameter space of
  baryogenesis in the $\nu$MSM}},  {\em JHEP} {\bf 07} (2019) 077,
  [\href{http://arxiv.org/abs/1808.10833}{{\tt arXiv:1808.10833}}].

\bibitem{Abada:2018oly}
A.~Abada, G.~Arcadi, V.~Domcke, M.~Drewes, J.~Klaric, and M.~Lucente, {\it
  {Low-scale leptogenesis with three heavy neutrinos}},  {\em JHEP} {\bf 01}
  (2019) 164, [\href{http://arxiv.org/abs/1810.12463}{{\tt arXiv:1810.12463}}].

\bibitem{Ghiglieri:2018wbs}
J.~Ghiglieri and M.~Laine, {\it {Precision study of GeV-scale resonant
  leptogenesis}},  {\em JHEP} {\bf 02} (2019) 014,
  [\href{http://arxiv.org/abs/1811.01971}{{\tt arXiv:1811.01971}}].

\bibitem{Klaric:2020lov}
J.~Klari\'c, M.~Shaposhnikov, and I.~Timiryasov, {\it {Uniting low-scale
  leptogeneses}},  \href{http://arxiv.org/abs/2008.13771}{{\tt
  arXiv:2008.13771}}.

\bibitem{Klaric:2021cpi}
J.~Klaric, M.~Shaposhnikov, and I.~Timiryasov, {\it {Reconciling resonant
  leptogenesis and baryogenesis via neutrino oscillations}},
  \href{http://arxiv.org/abs/2103.16545}{{\tt arXiv:2103.16545}}.

\bibitem{Drewes:2021nqr}
M.~Drewes, Y.~Georis, and J.~Klari\'c, {\it {Mapping the viable parameter space
  for testable leptogenesis}},  \href{http://arxiv.org/abs/2106.16226}{{\tt
  arXiv:2106.16226}}.

\bibitem{Dev:2014laa}
P.~Bhupal~Dev, P.~Millington, A.~Pilaftsis, and D.~Teresi, {\it {Flavour
  Covariant Transport Equations: an Application to Resonant Leptogenesis}},
  {\em Nucl. Phys. B} {\bf 886} (2014) 569--664,
  [\href{http://arxiv.org/abs/1404.1003}{{\tt arXiv:1404.1003}}].

\bibitem{Dev:2015wpa}
P.~S.~B. Dev, P.~Millington, A.~Pilaftsis, and D.~Teresi, {\it {Corrigendum to
  ''Flavour Covariant Transport Equations: an Application to Resonant
  Leptogenesis''}},  {\em Nucl. Phys. B} {\bf 897} (2015) 749--756,
  [\href{http://arxiv.org/abs/1504.07640}{{\tt arXiv:1504.07640}}].

\bibitem{Dev:2014wsa}
P.~Bhupal~Dev, P.~Millington, A.~Pilaftsis, and D.~Teresi, {\it
  {Kadanoff\textendash{}Baym approach to flavour mixing and oscillations in
  resonant leptogenesis}},  {\em Nucl. Phys. B} {\bf 891} (2015) 128--158,
  [\href{http://arxiv.org/abs/1410.6434}{{\tt arXiv:1410.6434}}].

\bibitem{Kartavtsev15}
A.~Kartavtsev, P.~Millington, and H.~Vogel, {\it {Lepton asymmetry from mixing
  and oscillations}},  {\em JHEP} {\bf 06} (2016) 066,
  [\href{http://arxiv.org/abs/1601.03086}{{\tt arXiv:1601.03086}}].

\bibitem{Fidler:2011yq}
C.~Fidler, M.~Herranen, K.~Kainulainen, and P.~M. Rahkila, {\it {Flavoured
  quantum Boltzmann equations from cQPA}},  {\em JHEP} {\bf 02} (2012) 065,
  [\href{http://arxiv.org/abs/1108.2309}{{\tt arXiv:1108.2309}}].

\bibitem{garbrecht11}
B.~Garbrecht and M.~Herranen, {\it {Effective Theory of Resonant Leptogenesis
  in the Closed-Time-Path Approach}},  {\em Nucl.Phys.} {\bf B861} (2012)
  17--52, [\href{http://arxiv.org/abs/hep-ph/1112.5954}{{\tt
  hep-ph/1112.5954}}].

\bibitem{garny11}
M.~Garny, A.~Kartavtsev, and A.~Hohenegger, {\it {Leptogenesis from first
  principles in the resonant regime}},  {\em Annals Phys.} {\bf 328} (2013)
  26--63, [\href{http://arxiv.org/abs/1112.6428}{{\tt arXiv:1112.6428}}].

\bibitem{Garbrecht:2014aga}
B.~Garbrecht, F.~Gautier, and J.~Klaric, {\it {Strong Washout Approximation to
  Resonant Leptogenesis}},  {\em JCAP} {\bf 09} (2014) 033,
  [\href{http://arxiv.org/abs/1406.4190}{{\tt arXiv:1406.4190}}].

\bibitem{Liu:1993ds}
J.~Liu and G.~Segr\`e, {\it {Unstable particle mixing and CP violation in weak
  decays}},  {\em Phys. Rev. D} {\bf 49} (1994) 1342--1349,
  [\href{http://arxiv.org/abs/hep-ph/9310248}{{\tt hep-ph/9310248}}].

\bibitem{covi96II}
L.~Covi and E.~Roulet, {\it {Baryogenesis from mixed particle decays}},  {\em
  Phys.Lett.} {\bf B399} (1997) 113--118,
  [\href{http://arxiv.org/abs/hep-ph/9611425}{{\tt hep-ph/9611425}}].

\bibitem{Racker20}
J.~Racker, {\it {CP violation in mixing and oscillations in a toy model for
  leptogenesis with quasi-degenerate neutrinos}},  {\em JHEP} {\bf 04} (2021)
  290, [\href{http://arxiv.org/abs/2012.05354}{{\tt arXiv:2012.05354}}].

\bibitem{Jukkala:2021cys}
H.~Jukkala, K.~Kainulainen, and P.~M. Rahkila, {\it {Flavour mixing transport
  theory and resonant leptogenesis}},
  \href{http://arxiv.org/abs/2104.03998}{{\tt arXiv:2104.03998}}.

\bibitem{Beuthe01}
M.~Beuthe, {\it {Oscillations of neutrinos and mesons in quantum field
  theory}},  {\em Phys. Rept.} {\bf 375} (2003) 105--218,
  [\href{http://arxiv.org/abs/hep-ph/0109119}{{\tt hep-ph/0109119}}].

\bibitem{fuchs16}
E.~Fuchs and G.~Weiglein, {\it {Breit-Wigner approximation for propagators of
  mixed unstable states}},  {\em JHEP} {\bf 09} (2017) 079,
  [\href{http://arxiv.org/abs/1610.06193}{{\tt arXiv:1610.06193}}].

\bibitem{1963AnPhy22}
R.~G. {Sachs}, {\it {Interference phenomena of neutral K mesons}},  {\em Annals
  of Physics} {\bf 22} (May, 1963) 239--262.

\bibitem{PhysRevD.48.4310}
C.~Giunti, C.~W. Kim, J.~A. Lee, and U.~W. Lee, {\it Treatment of neutrino
  oscillations without resort to weak eigenstates},  {\em Phys. Rev. D} {\bf
  48} (Nov, 1993) 4310--4317.

\bibitem{roulet97}
E.~Roulet, L.~Covi, and F.~Vissani, {\it {On the CP asymmetries in Majorana
  neutrino decays}},  {\em Phys. Lett.} {\bf B424} (1998) 101--105,
  [\href{http://arxiv.org/abs/hep-ph/9712468}{{\tt hep-ph/9712468}}].

\end{thebibliography}\endgroup

\end{document}